\documentclass{elsart}
\usepackage{graphicx}
\begin{document}
\runauthor{Hashibon, Adler, Finnis, and Kaplan}
\begin{frontmatter}
\title{Atomistic Study of Structural Correlations at a Liquid-Solid Interface
}
\author[Rome]{Adham Hashibon\corauthref{cor1}}
\author[Rome]{Joan Adler}
\author[Paestum]{Michael W. Finnis}
\author[Baiae]{Wayne D. Kaplan}
\corauth[cor1]{phadham@technion.ac.il, Tel:+972-4-8292043, Fax: +972-4-8221514}
\address[Rome]{Department of Physics, Technion, Haifa 32000, Israel}
\address[Paestum]{Atomistic Simulation Group, School of Mathematics and Physics, \\ {\em  The Queen's University of Belfast, Belfast BT7 1NN}}
\address[Baiae]{Department of Materials Engineering, Technion, Haifa 32000, Israel}
\begin{abstract}
  Structural correlations at a liquid-solid interface were explored
  with molecular dynamics simulations of a model aluminium system
  using the Ercolessi-Adams potential and up to 4320 atoms.  Substrate
  atoms were pinned to their equilibrium crystalline positions while
  liquid atoms were free to move.  The density profile at the
  interface was investigated for different substrate crystallographic
  orientations and temperatures. An exponential decay of the density
  profile was observed, $\rho (z) \sim \rm{e}^{- \kappa z}$, leading
  to the definition of $\kappa$ as a quantitative measure of the
  ordering at the liquid solid interface. A direct correlation between
  the amount of ordering in the liquid phase and the underlying
  substrate orientation was found.
\end{abstract}
\begin{keyword}
solid-liquid interfaces, computer simulation,  aluminium, layering
\PACS 68.45.-v, 68.35.-p, 02.70.Ns,  68.45.Gd
\end{keyword}
\end{frontmatter}
\newcommand{\x}{\times}
\newcommand\mean[1]{{\left\langle#1\right\rangle}}
\section{Introduction}
Metal-ceramic interfaces play a prominent role in a variety of
technological applications and processes that range from electronic
devices, protective coatings, high-temperature structural components,
and liquid phase joining processes. The functionality of these systems
depends crucially on their macroscopic properties such as fracture
strength, yield, and electrical conductivity. These properties are
strongly correlated with microscopic details of the metal-ceramic
interface such as wetting, chemistry, diffusion, and structure.
Correlating macroscopic properties to the structure and chemistry of
interfaces is one of the most intriguing topics in materials science.
Experimental studies of the atomistic structure of a solid-{\it
  liquid} internal interface are technically difficult to
conduct~\cite{Huisman1997,Reichert2000}. Therefore, atomistic
simulations of metal ceramic interfaces can serve as an important tool
to understand and predict the effect of the interface region on the
material properties.  {\it Ab-initio} electronic structure
calculations provide detailed information of the chemical bonding
across the interface, but are very expensive in terms of computer
power and time. Consequently, they are limited to modeling small
systems, which are not of sufficient size to contain large structural
defects.  Moreover, a {\it dynamical} simulation of a liquid-solid
interface is beyond both system size and CPU limitations for {\it
  ab-initio} calculations.

Atomistic simulations, such as Molecular Dynamics (MD) or Monte Carlo
permit the controlled study of these systems at the atomistic level
for a large number of atoms and for large structures. However the main
limitation to such simulations is the lack of appropriate interatomic
potential schemes which can model both metallic and ionic bonding
across the interface. Nevertheless, simplified models can be used to
obtain qualitative basic insights into the problem.  In this study we
introduce a model system in which the ceramic is assumed to be
composed of atoms pinned to their equilibrium lattice positions, while
the metal atoms are free to evolve under the influence of their
interatomic potential.

The atoms of a liquid metal which are adjacent to a rigid crystalline
substrate are in an environment which is strongly affected by the
symmetry of the underlying substrate. Theoretical
studies~\cite{Howe1996}, which are mainly computational, have shown
that ordering occurs in the first layers of the liquid adjacent to the
crystal surface.  The same result emerges from experimental studies of
solid-liquid interfaces~\cite{Huisman1997,Howe1996}. A particularly
interesting study with respect to the current project was conducted by
Huisman~\etal\cite{Huisman1997}, who investigated the interface
structure of liquid gallium in contact with a diamond $(111)$ surface.
They observed pronounced layering of the liquid metal density profile
which decays exponentially with increasing distance from the diamond.
Moreover, the interlayer spacing was equal to the repeat distance of
$(100)$ planes of upright gallium dimers in solid $\alpha$-Ga. Thus it
appears that the liquid near the surface assumes a solid-like
structure similar to the $\alpha$-phase.

Consequently, at a solid-liquid interface, there should be a
transition from a solid-like phase in the liquid near the interface to
a liquid like phase further away from the interface.  The extent over
which such a transition occurs should be directly related to the
structure of the underlying surface. In the current research,
simulations have been conducted to study the density profile and
structure of the liquid-metal/hard-wall interface as a function of
temperature and substrate structure. In particular we show that the
decay of the density profile is quantitatively and qualitatively
related to the underlying structure of the substrate.
 

\section{System and Simulation Method}
\label{sec:methods}

\begin{figure}
\centering\includegraphics[width=10.0cm]{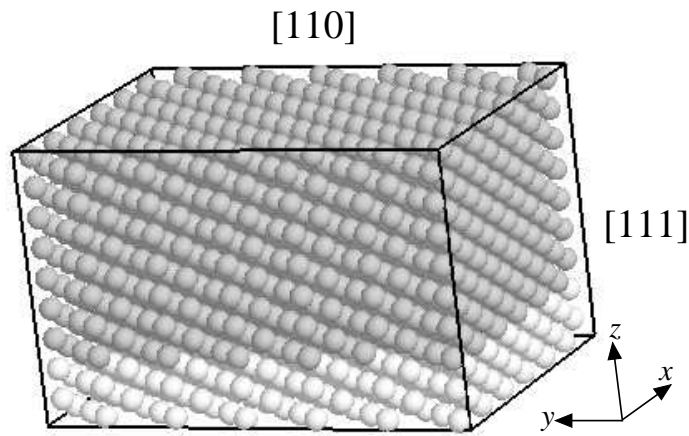}

\caption{An example of a cell for the simulation of a system with a
free {(}110{)} surface.  The total number of atoms is \protect\(
1944\protect \), out of which \protect\( 432\protect \) atoms are
'solid' (light gray) and the rest are ``liquid'' (dark gray) aluminium
atoms.}
\label{fig:fig1}
\end{figure}

The initial configuration for the simulations was a crystalline face-centered-cubic (fcc) aluminium lattice. A liquid-solid interface was formed by pinning the atoms in the first few layers to their ideal lattice positions, and allowing the rest of the atoms to move freely under the effect of the interatomic potential.  The lattice parameter of the fcc aluminium crystal was $a=4.1$\AA, which is close to the value obtained for aluminium at the melting point ($T=940$K) with the potential model used here~\cite{EA1994} (see below for more details of the potential).  Three orientations for the terminating substrate plane were used: $(100)$, $(110)$, and $(111)$.  An example of a simulation cell with a $(110)$ interface is shown in Fig.~\ref{fig:fig1}.  The light gray atoms were fixed to their equilibrium positions throughout the simulation. The number of particles in each sample, as well as the geometry, are described in Table~\ref{tab:geometry}.

In the plane of the interface (the $xy$ plane) periodic boundary conditions were applied.  In the direction perpendicular to the interface (the $z$ direction), the boundary conditions are expected to simulate the bulk media on either side of the interface.  On the rigid (substrate) side where atoms are fixed to crystalline positions, it is sufficient to require that the extent of the region in the $z$ direction to be larger than the cutoff of the interaction potential. In this way liquid atoms near the interface do not ``see'' the bottom of the rigid layer, hence it acts like a semi infinite bulk system.  On the liquid (metal) side the situation is more difficult. If periodic boundary conditions are applied, the first rigid layer (see Fig.~\ref{fig:fig1}) and the last liquid layer will interact forming a second interface in the system. These two interfaces can then interact with each other, unless the dimension in the z direction is very large so that the interaction is negligible. In addition, due to the confinement of the simulation cell, the system will not be able to respond to volume changes caused by stress at the interface.  One way to overcome this difficulty is to use a model in which the sample is enclosed by two moving rigid walls as proposed by Lutsko {\it et al.}~\cite{Lutsoko88}. Here, to allow for volume deformations, the $z$ dimension is considered as a dynamic variable and a formulation is introduced to allow a flexible length of the sample normal to the interface in response to stresses in the simulation cell. 

In this study a liquid layer is deposited above the solid layer, and then a vacuum region is inserted with periodic boundary conditions in all directions. In this way we have one free liquid interface and one internal solid-liquid interface. Provided that the height of the liquid layer is large enough in the $z$ direction, there will be no interaction between the free liquid surface and the internal rigid-liquid interface. In addition the system will be free to respond to stresses at the interface since there is nothing to limit the liquid from expanding in the $z$ direction. 

Since periodic boundary conditions were also applied in the z direction, there are two requirements on the extent of the vacuum region.  Firstly, it should be larger than the cutoff distance of the interatomic potential, otherwise the upper most liquid layer will interact with the bottom of the rigid layer. Secondly, if liquid atoms evaporate from the free surface and enter into the vacuum region, there is a chance that these vapor atoms would reach the end of the vacuum and interact with the bottom region via the periodic boundary condition. However, as discussed below, there are practically no vapor atoms during the simulation time. Hence it is sufficient to have a large enough vacuum based on the cutoff distance only. In practice we chose a vacuum region of 15 times the extent of the cutoff distance. 
The system was simulated using an MD technique, which consists of the numerical integration of Newton's equation of motion for the various atoms~\cite{FS1996}. The velocity-Verlet integration algorithm~\cite{FS1996,AT1987} was used in the simulations.  For the interatomic potential, an embedded atom potential developed by Ercolessi and Adams (EA)~\cite{glue} was used. This potential was constructed by the so called force matching method, whereby the potential was fitted to a very large amount of data obtained from both experiment and first principle calculations, with emphasis given to match the interatomic forces obtained from the potential to those obtained from first principles. The potential has been tested in detail for aluminium\cite{EA1994,Hansen99}, and was found to be consistent with experimental results. As an example, the calculated melting point for aluminium is T=939$\pm$5K in excellent agreement with the experimental value of T=933.6K. 

One advantage for using a more realistic metallic potential (instead of a simpler pair potential such as a Leonard-Jones (LJ) 6-12 potential), is that it gives the realistic low vapor pressure characteristic of liquid metals.  Simulations that were performed with a LJ potential resulted in evaporation of a large portion of the liquid.  In the case of liquid aluminium, the  low vapor pressure results in as little as $10^{-8}$  vapor atoms in a simulation box of cubic side equal to $\sim 40$\AA, therefore, in effect no vapor atoms are observed during the simulations.

As a validation of our implementation of the EA potential we calculated the calorie curve for a bulk liquid system i.e. the internal energy as a function of temperature $E(T)$, the pressure, and the translational order parameter as functions of temperature.  In all cases we observed a discontinuous jump at a temperature of $T=930\pm 15$K, which is consistent with the calculated melting temperature of Ercolessi and Adams\cite{glue}. Of course this is not the best way to measure the melting point, but it suffices as a preliminary check of our correct implementation of the EA potential in our code scheme. 

As stated above, the size of the substrate (the rigid layers) is determined by the cutoff of the interaction potential. The range of effective interactions in an EAM type of potential is actually twice the cutoff radius of the glue part, which for the EA potential is $R_{cutoff} = 5.558$\AA. For example, with a lattice constant of $a=4.1$\AA, and a (220) inter-planar distance (or d-spacing) of $d_{220}=a/\sqrt{8}$, at least eight (220) atomic planes should be introduced into the rigid region~\footnote{ Normally the d-spacing of a family of planes with Miller indices (hkl) is given by: $d_{hkl} = a_{latt}/\sqrt{h^2 + k^2 + l^2}$. For example in the case of (110) planes: $d_{110} = a_{latt}/\sqrt{2}$. However, this is not the smallest distance between two consecutive planes having the same in-plane structure as the (110) plane. The family of planes that includes all planes with a (110) structure is in fact the (220) family. Hence in calculating the d-spacing for the (110) surface, we used $d_{220}$. Similarly for the (100) family, we used $d_{200}$.}. For the [111] direction, with an inter-planar distance of $d_{111}=a/\sqrt{3}$, five planes are required. In this case, the ``liquid'' atoms do not interact with the bottom of the rigid sample, which is effectively a semi-infinite bulk. A simple way of modifying the substrate liquid interaction is to reduce the number of substrate planes from this value, and we have included such results here.

The atoms in the fixed region were excluded from the equations of motion, although the forces they exert on the adjacent layers {\it were} included. In this way these fixed layers can be thought of as being a part of a different material with a much higher melting temperature, such as a ceramic.

The temperature of the simulations was controlled by a simple \textit{ad-hoc} rescaling of the velocities of the particles, so that the required average kinetic temperature was reached~\cite{AT1987}. Two schemes of temperature control were used; rescaling of all particle velocities, and rescaling only of the velocities of atoms adjacent to the fixed region, i.e. of the two layers next to the fixed one. The time step of the MD integration in normalized units was $t=0.02$, where the normalized unit is $\tau = 4.25\times10^{-14}$ in real seconds. 

\subsection{Computing the Density Profile}
\label{calculation of properties}


The density profile $\rho(z)$ is defined as the average density of particles in a slice of width $\Delta z$ parallel to the hard wall surface and centered around $x$. The simulation cell is divided into equal layers or bins parallel to the interface.  The expression for the density profile is 
\begin{equation}
\label{eq:rho}
\rho(z) = \frac{\mean{N_z}}{L_xL_y\Delta z} \nonumber
\label{eq:N_x}
\end{equation}
where $L_x$ and $L_y$ are the $x$ and $y$ dimensions of the cell, respectively, and $z$ is perpendicular to the interface, $\Delta z$ is the bin width, and $N_z$ is the number of particles between $z-\Delta z/2$ and $z+\Delta z /2 $ at time $t$. The angled brackets indicate a time average.

In order to reduce the statistical error of the sampling, a proper choice of bin width must be made. Very small bin widths results in too few particles at each time step, hence a large scatter of the data. Very large bins will not show the actual dependence of the density profile over distance.  Two basic width scales have been used: a coarse scale, in which the width of the bins was set equal to the bulk crystal d-spacing for a particular orientation, and a fine scale for which each coarse scale bin was divided into 10 or 25 sections. 

\subsection{Equilibration and Computation of  Averages}
\label{sec:equilibriation}

\begin{figure}
\vspace{4.0cm}
\centering{\includegraphics[width=10.0cm]{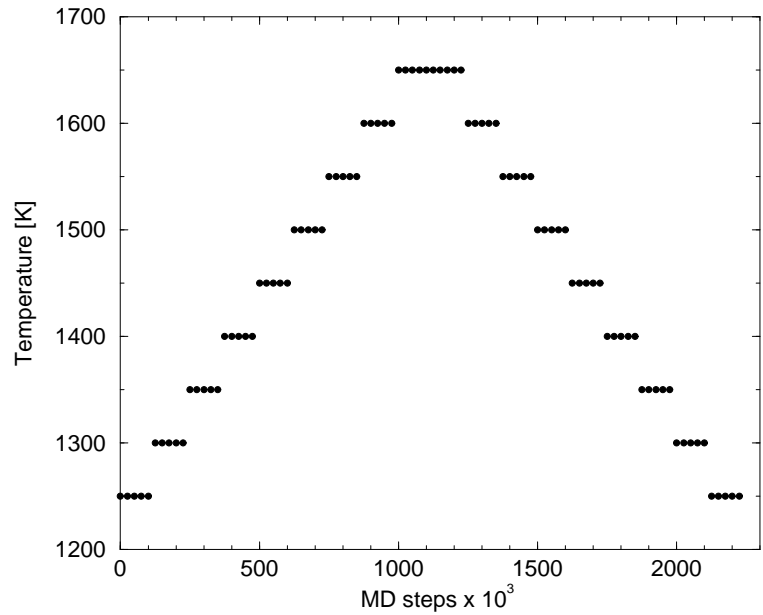}}
\caption{An example of an annealing schedule for system number 5 (see
Table~\ref{tab:geometry}). The state of the system at the final
(target) temperature is compared with the initial state. The schedule is repeated until the system is in equilibrium, that is when the
 energy and density profile in the final state and initial state are
the same.}
\label{fig:anealing}
\end{figure}

For each system described in Table~\ref{tab:geometry} a series of simulations was conducted at temperatures ranging from $T=800$K up to $T=2000$K in steps of $50$K. To obtain an equilibrated sample at a particular temperature, an ideal configuration is annealed by heating the system in steps of 50K to a higher temperature than required, and then cooling down in steps of 50K to the target temperature. Fig.~\ref{fig:anealing} shows  one such annealing schedule for system No. 1 (see Table~\ref{tab:geometry}). After annealing, the system is further equilibrated at the target temperature for at least $50,000$ MD steps for the smaller systems and $100,000$ steps for the larger systems, before sampling begins. 

Measurements of the density profile are obtained by accumulating $\rho(z)$ via equation~(\ref{eq:rho}) over 20,000 to 50,000 MD steps, then averaging over these to produce a single measurement. In equilibrium, these measurements of $\rho(z)$ do not change substantially in time, and are independent of the annealing schedule.


\section{Results and Discussion}
\label{sec:results}

\begin{figure}
\centering 
\includegraphics[width=12.0cm]{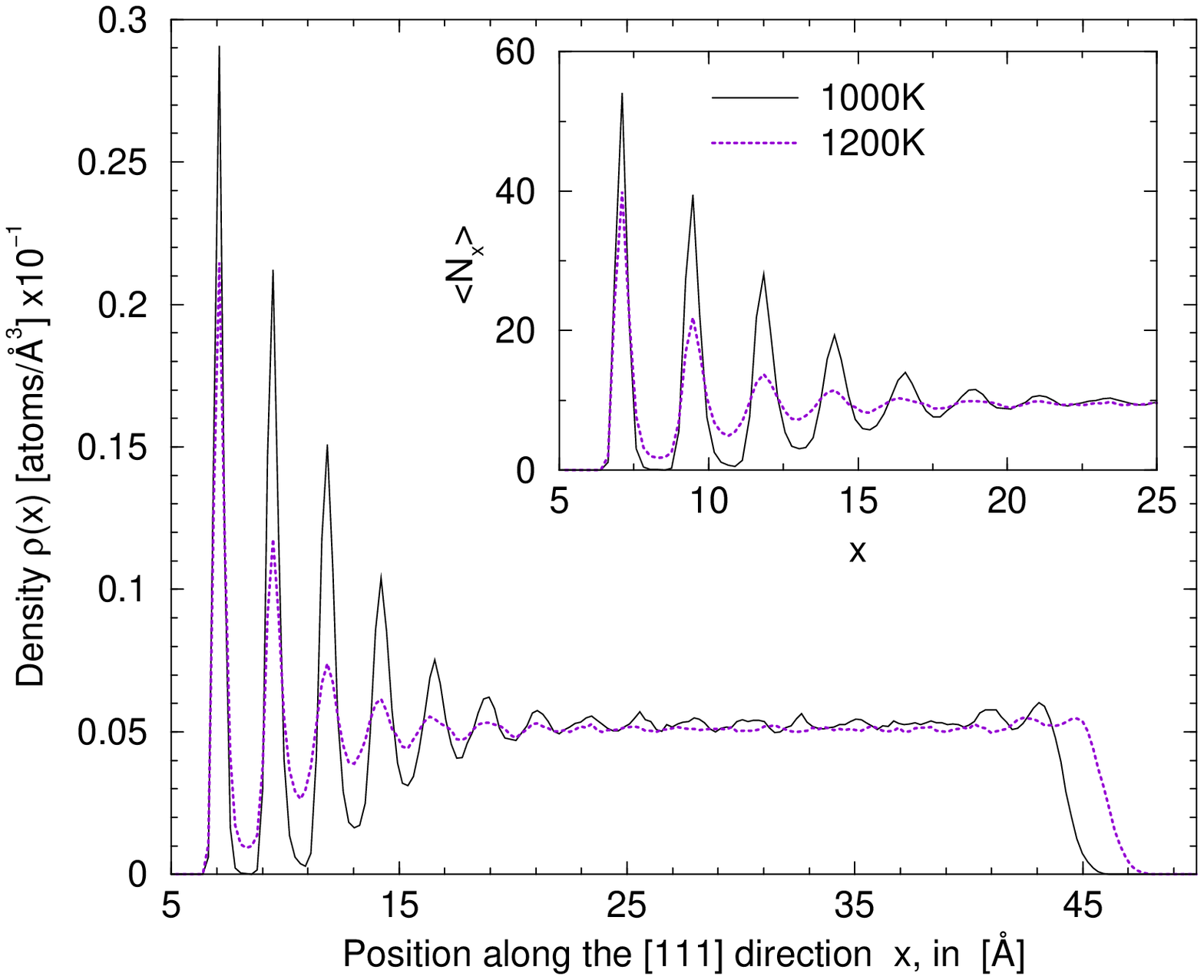}
\caption{The density profile $\rho(z)$ as a function of distance from the hard wall $z$, is  shown for a (111) system at two temperatures: $T_1=1000K$, and $T_2 = 1200K$. The insert is a plot of the average number of atoms per bin $\mean{N_x}$ for the region near to the rigid wall. Data is from a simulation on system number 4 described in Table~\ref{tab:geometry}}
\label{fig:denprof-compare111}
\end{figure}
\begin{figure}
\centering 
\includegraphics[width=12.0cm]{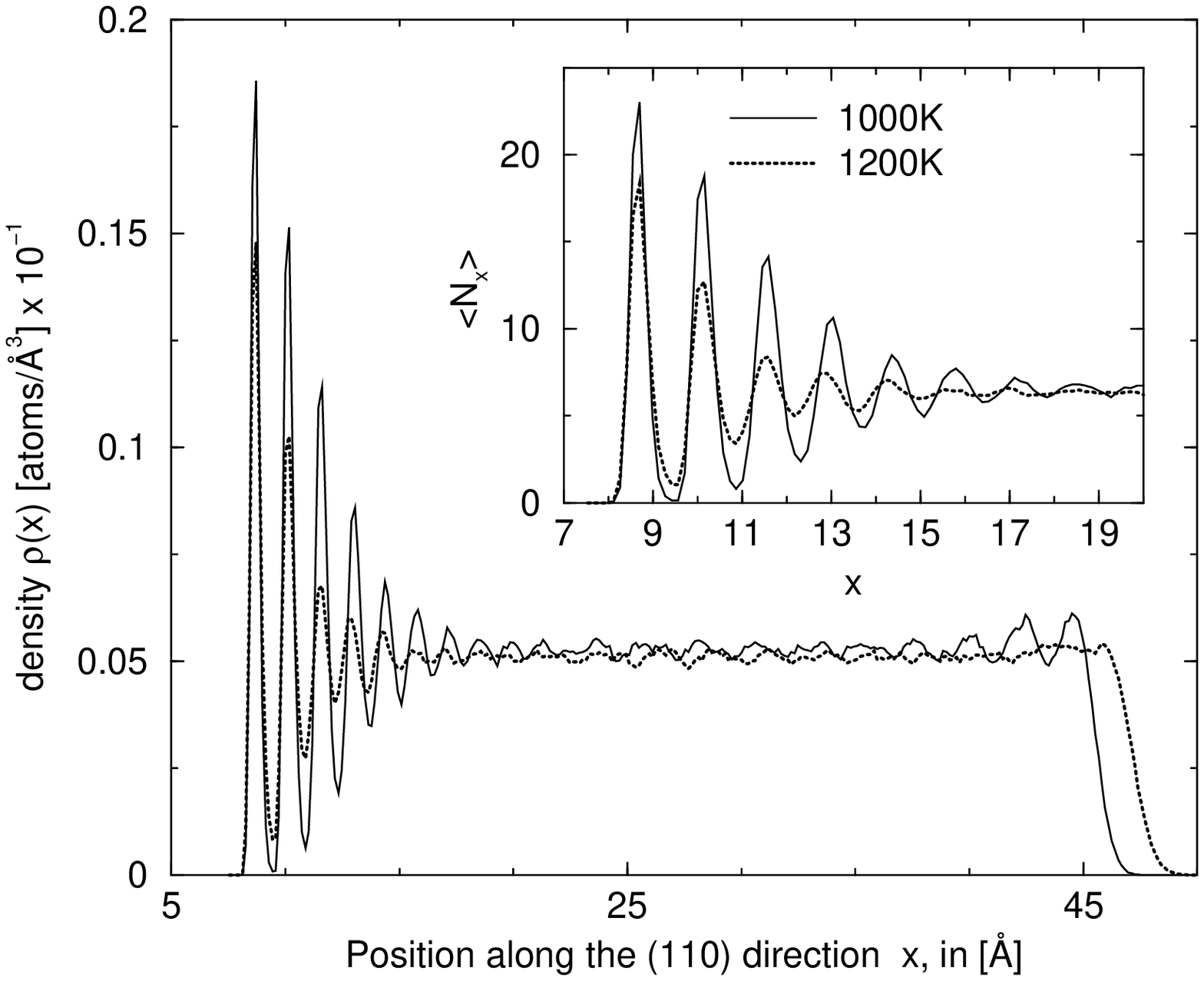}
\caption{The density profile $\rho(z)$ as a function of distance from the hard wall $z$, is  shown for a (110) system at two temperatures: $T_1=1000K$, and $T_2 = 1200K$. The insert is a plot of the average number of atoms per bin $\mean{N_x}$ for the region near to the rigid wall. Data is from a simulation on system number 2 described in Table~\ref{tab:geometry}.}
\label{fig:denprof-compare110}
\end{figure}
\begin{figure}
\centering 
\includegraphics[width=12.0cm]{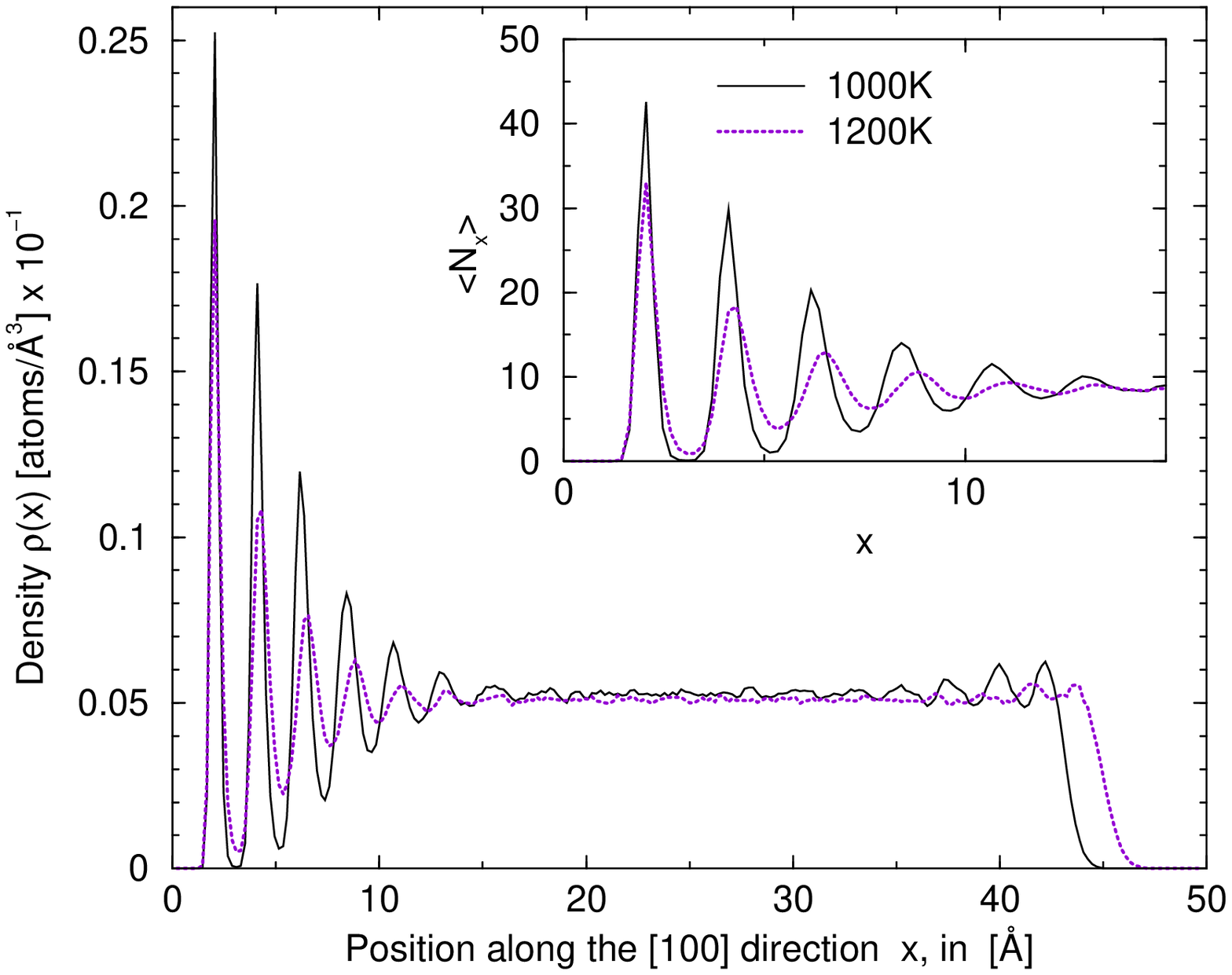}
\caption{The density profile $\rho(z)$ as a function of distance from the hard wall $z$, is  shown for a (100) system at two temperatures: $T_1=1000K$, and $T_2 = 1200K$. The insert is a plot of the average number of atoms per bin $\mean{N_x}$ for the region near to the rigid wall. Data is from a simulation on system number 1 described in Table~\ref{tab:geometry}.}
\label{fig:denprof-compare100}
\end{figure}

The density profiles $\rho(z)$ of the liquid part of the samples are shown at two temperatures in Figs.~\ref{fig:denprof-compare111}, \ref{fig:denprof-compare110}, and~\ref{fig:denprof-compare100} for the (111), (110), and (100) interfaces, respectively. The inserts in the figures are enlargements of the profiles of the first few layers, with the abscissa given as the average number of particles per bin $\mean{N_z}$ [see equation~(\ref{eq:N_x})], rather than as the density.  These density profiles show large oscillations corresponding to the layering of the liquid near the hard wall.  The oscillations dampen gradually within the interfacial region, until the uniform density of the liquid phase is reached, which is the same in all systems: $\rho_{l} = 0.0051 \pm 0.0005 \; \rm{atoms/\AA^3}$. As expected, the density profile at a higher temperature decays faster as a function of the distance from the interface than the density profile at a low temperature.

\begin{figure}
\centering 
\includegraphics[width=12.0cm]{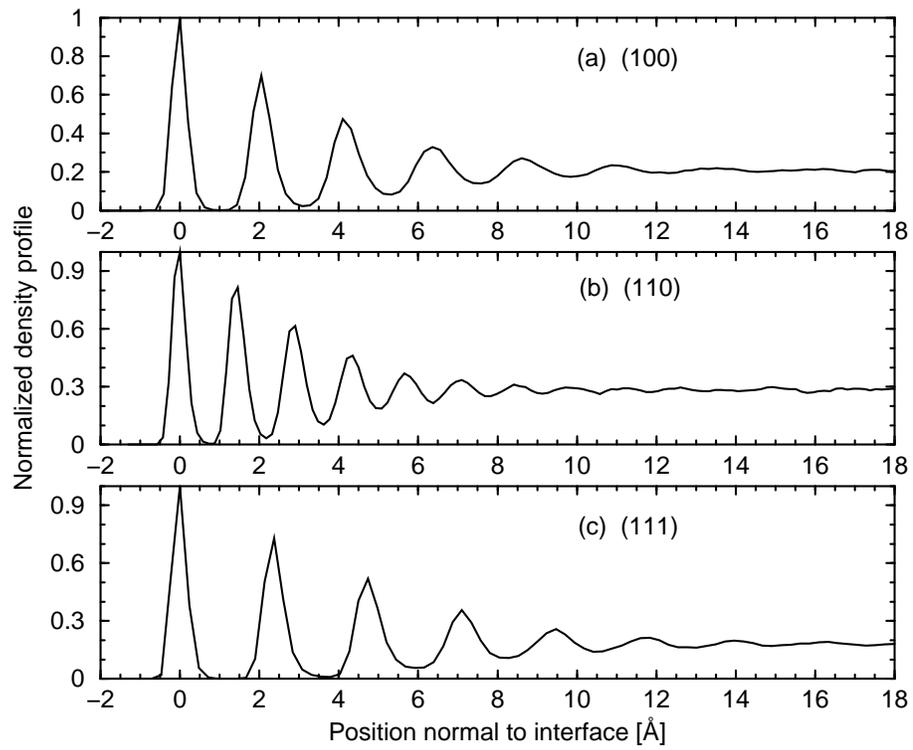}
\caption{Normalized fine scale density profile for (a): (111), (b):(110), and (c) (100) interfaces at the same temperature (1000K).}
\label{fig:denprof-compare2}
\end{figure}

\begin{figure}
\centering 
\includegraphics[width=12.0cm]{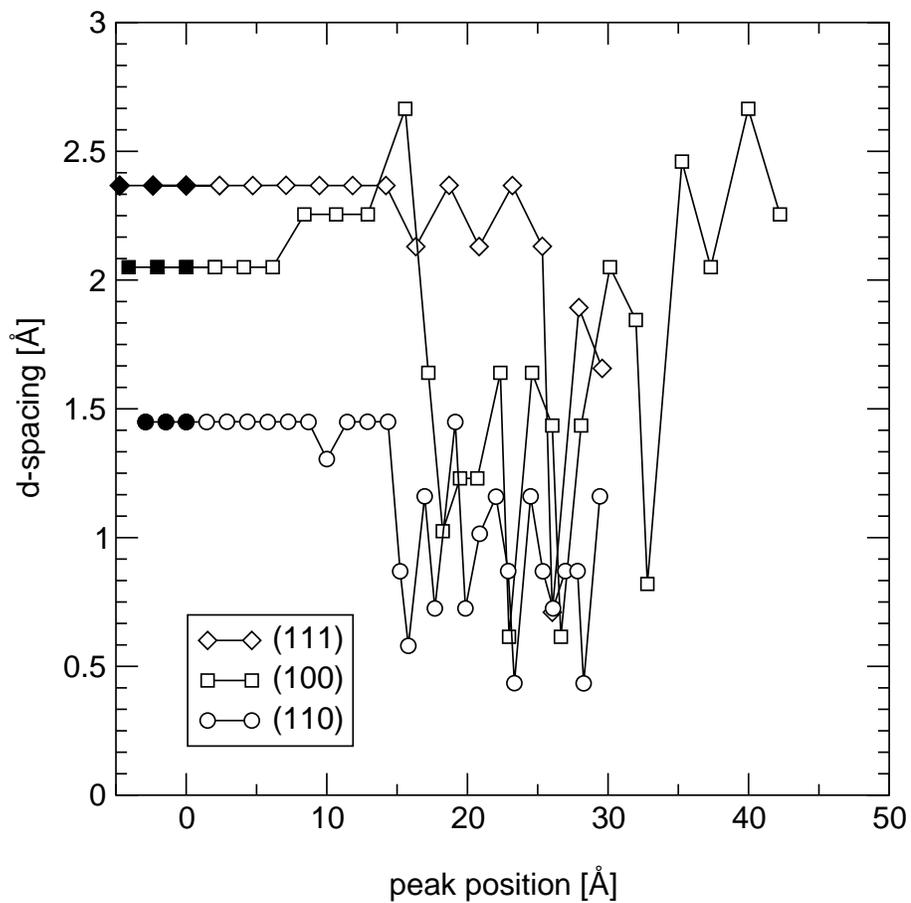}
\caption{Layer separation across the samples at T=1000K,  the dark symbols are within the
rigid substrate, and the open symbols are within the liquid.  }
\label{fig:interlayer1}
\end{figure}

In order to facilitate a direct comparison between the density profiles, the density is normalized by the magnitude of the first peak as shown in Fig.~\ref{fig:denprof-compare2}, where the density profiles for these systems are plotted at $T=1000$K. It is easy to see from Fig.~\ref{fig:denprof-compare2} that ordering at the (111) and (100) interfaces extends further into the liquid than at the (110) interface. However, the number of density peaks is very similar in all directions, while the distance between consecutive peaks varies substantially with interface direction.  This distance, or interlayer spacing can be identified with the d-spacing of the quasi solid region within the liquid.  The calculation of the d-spacing is made with an algorithm that searches for maxima in the density profile, and then calculates the difference in the locations of each consecutive maxima, giving the d-spacing. The d-spacing is plotted in Fig.~\ref{fig:interlayer1} at $T=1000$K. The data in Fig.~\ref{fig:interlayer1} includes the analysis of the d-spacing even well inside the liquid region, where the "maxima" are in random locations and are due to noise in the data. This explains the large scatter of data at a distance from the interface, and can serve as an indication of the point at which the ordering terminates. For example, it is observed that the ordering of the (111) interface extends up to almost 25\AA, before becoming too noisy.  It is also clear that the interlayer spacing in the quasi-liquid is the same as that in the rigid substrate, at least in the first few layers. However, the (100) layers show a very large expansion of the d-spacing in the direction of that of the (111) system.  Similar behavior was previously observed~\cite{Ruslan98,BG-V,Tomagnini96}, and arises from the fact that perturbations within the liquid are more energetically favored if their periodicity is $2\pi/Q_0$, where $Q_0$ is the wave vector at the first peak of the structure factor $S(Q)$. For an fcc liquid metal this distance is very close to $d_{111}$~\cite{Tomagnini96}. The d-spacing starts to deviate from the bulk value far away from the substrate, as can be seen for the (111) case, where $d$ alternates between two values very close to $d_{111}$ before becoming completely random. For the (110) case a shift in the interlayer spacing is not observed. This may be attributed to the large difference between $d_{110}$ and $d_{111}$.

\begin{figure}
\centering
\begin{tabular}{c}
{\large \bf (a)} \\
\includegraphics[width=10.0cm]{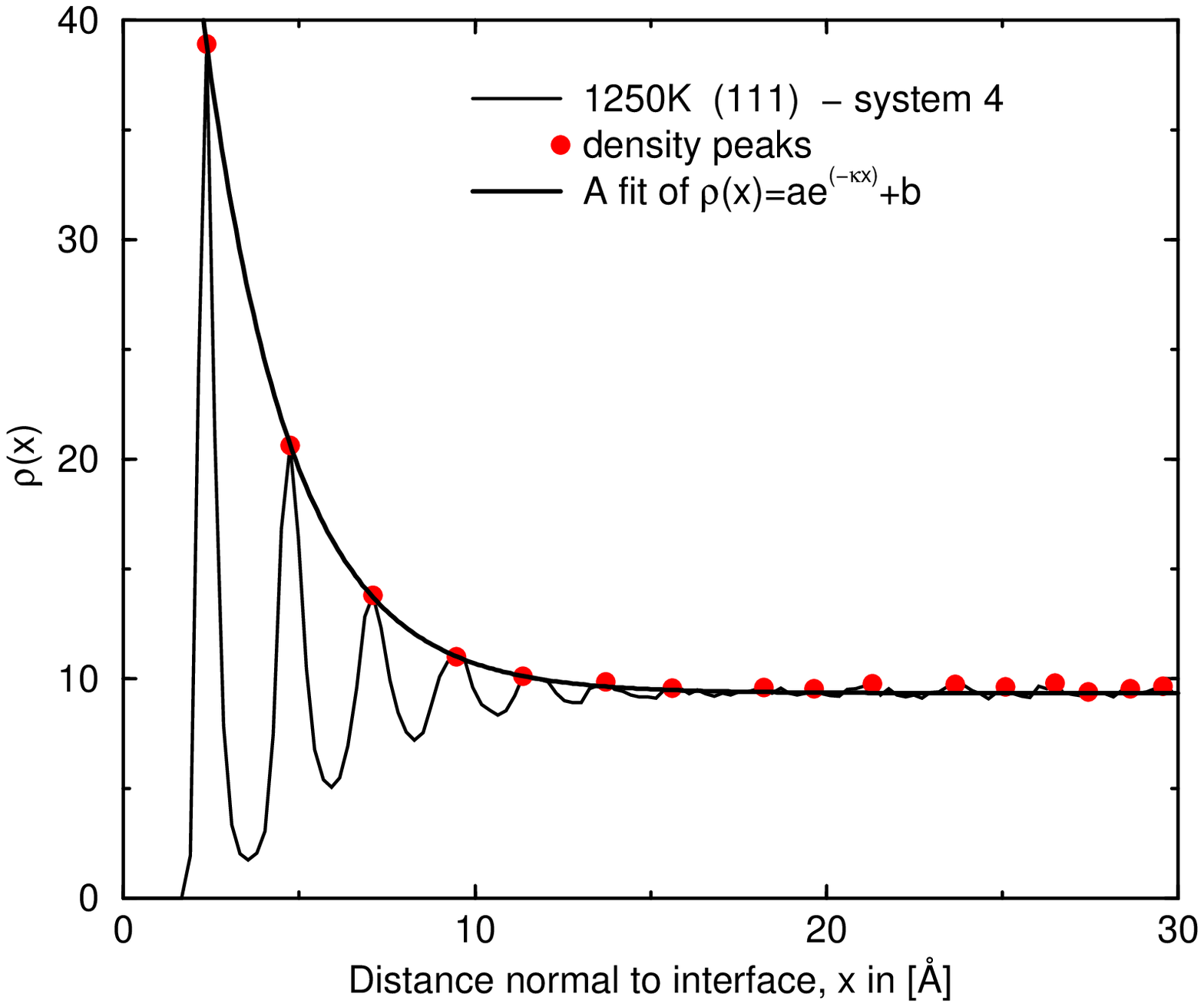} \\
{\large \bf (b)} \\
\includegraphics[width=10.0cm]{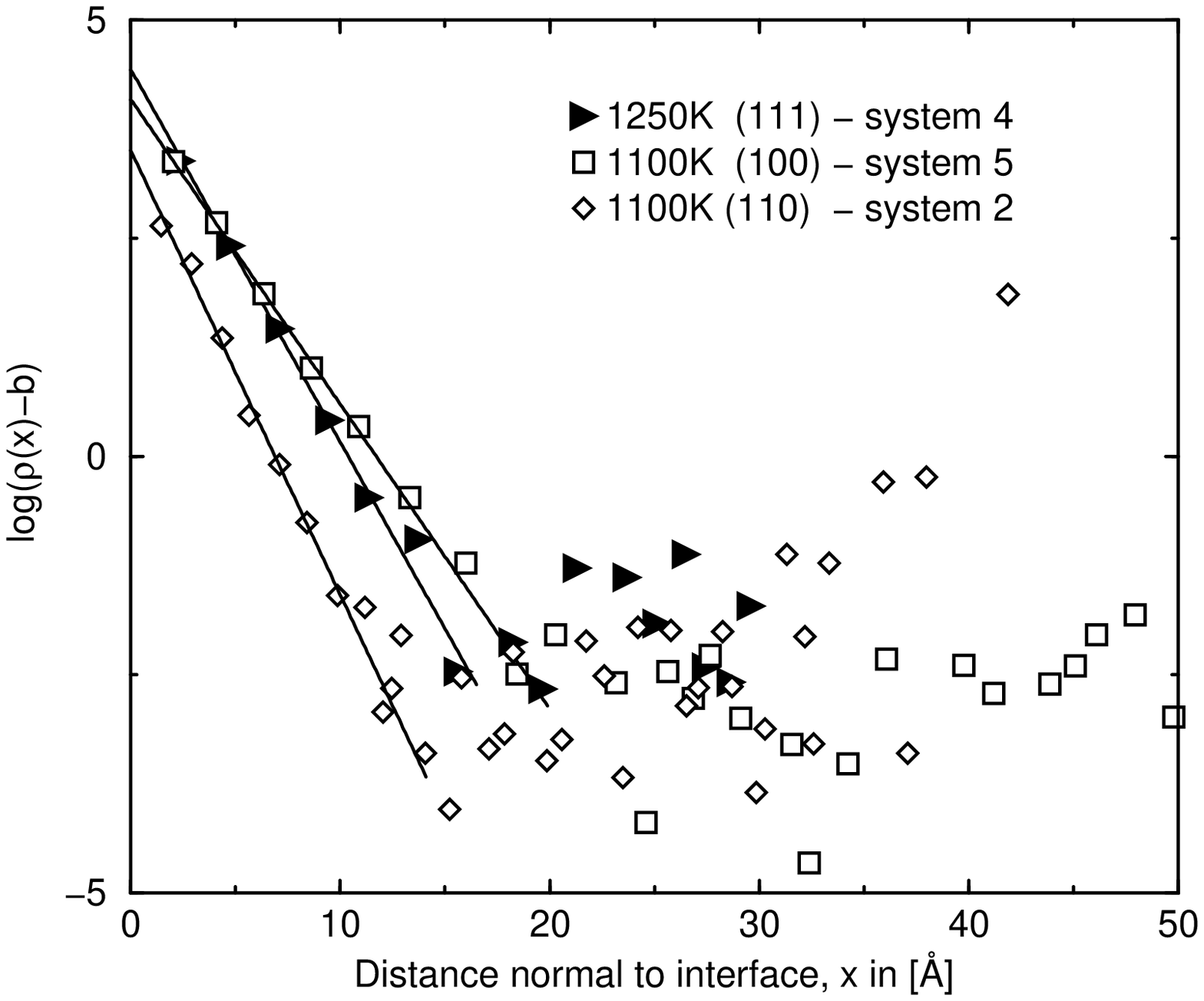}
\end{tabular}
\caption{The envelope of the density profile obtained for a (110)
sample at $T=1000K$. System numbers are defined in Table~\ref{tab:geometry}.}
\label{fig:fitexample}
\end{figure}

In order to analyze the decay of the density profile as a function of temperature and substrate orientation, the envelope of the density profile was extracted and plotted against distance from the surface, as shown in Fig.~\ref{fig:fitexample} (a). The profile is then fitted assuming an exponential decay: $\rho(z) \sim \exp{(-\kappa z)}$, where $\kappa = 1/\xi$, and $\xi$ is the correlation length at the interface.  The parameter $\kappa$ can then be used to {\em quantitatively} describe the amount of {\em disorder} at the interface. The exact function that has been used in the fitting has an extra constant term, $b$, to account for the background liquid density, and a normalization factor, $a$: 
\begin{equation}
\rho(z) = a\rm{e}^{-\kappa z} + b
\label{eq:denprof}
\end{equation}
This form of decay of the density profile is typically obtained from a mean field treatment of binary fluid interfaces~\cite{Tomagnini96,tarazona}. A good check of the validity of this assumption is that ${\rm log}(\rho(z)-b)$ is linear in $z$. A few typical examples are shown in Fig.~\ref{fig:fitexample} (b). The large scatter for points far from the solid-liquid interface, i.e. well within the normal liquid phase, is due to the sensitivity of the logarithm to small arguments, since the difference between $\rho(z)$ and $b$ for points inside the liquid is very small.

\begin{figure}
\centering
\includegraphics[width=12.0cm]{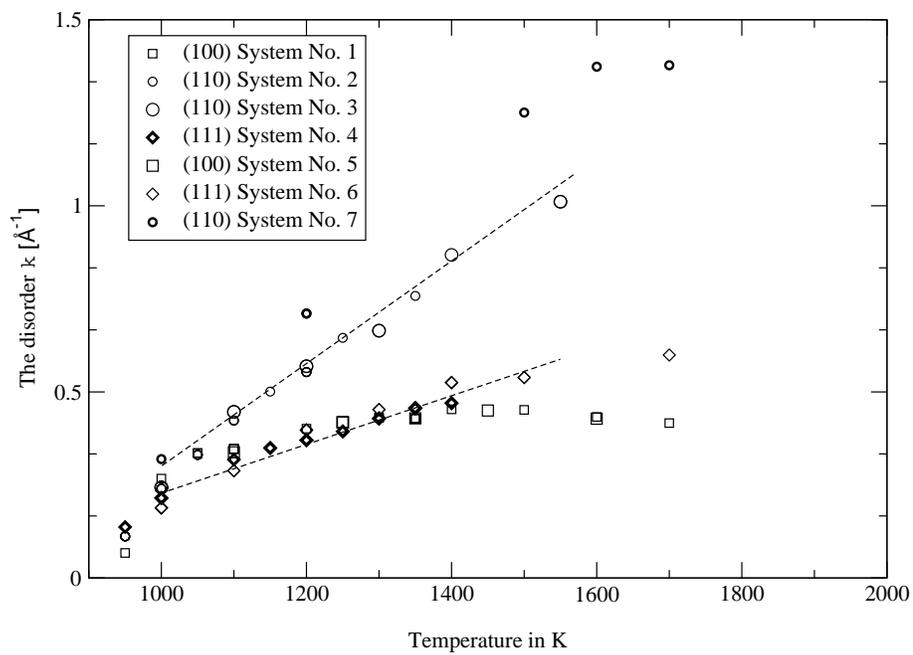}
\caption{The disorder parameter $\kappa$ vs. temperature for the three surfaces.}
\label{fig:kappa3faces}
\end{figure}

The results of the density profile analysis for the three interfaces, namely (111), (110) and (100) are shown in Fig.~\ref{fig:kappa3faces}. Here the disorder parameter $\kappa$ is plotted against temperature for the three orientations (111), (110) and (100), for various systems as indicated in the figure.  It is clear that the finite size effects are very small, as can be seen from the compatible $\kappa$ values for systems No. 1 and 5 (see Table~\ref{tab:geometry}) and systems No. 2 and 3.  Moreover, we note that the width of the underlying rigid substrate does not alter the qualitative behavior of $\kappa$, as seen from the results for systems No. 6 and 7.  However, there is a tendency for $\kappa$ to be larger for a thick substrate, as evident mainly from the results for system 7.  This can be attributed to the ``gluing'' nature of the potential, namely that the energy of an atom is lower when it has more neighbors. Hence when there are more substrate layers the energy of the overall system is lowered, and the net force exerted on the liquid is lower than with a thinner substrate which leads to less ordering in the liquid phase. 
        
The amount of disorder, $\kappa$, is consistently smaller for the (111) and (100) surfaces than for the (110) system, as shown in Fig.~\ref{fig:kappa3faces}. In other words, the ordering at the (111) and (100) interfaces is larger, and extends further into the liquid, than the ordering at the (110) interface. 

The close-packed surfaces (100) and (111) in an ideal system have a lower energy than the (110) surface, and therefore are more stable against thermal effects.  The formation energies of an {\em ideal surface}, as calculated with the EA potential~\cite{Hansen99}, and with LDA~\cite{Stumpf} are shown in table~\ref{tab:energy}. The formation energies of an ideal (111) and (100) surfaces are similar, with the difference between them being  only $\sim 0.1$ eV/atom with the EA potential. The formation energy of the (110) surface is $0.26$ eV/atom higher than that of the (100) surface. The similarity of the formation energies of the ideal (111) and (100) systems is compatible with the behavior of $\kappa$ in Fig.~\ref{fig:kappa3faces}, where $\kappa$ for the (111) and (100) systems are almost indistinguishable. This suggest that the the energy balance between the different surfaces is not altered by the presence of a liquid phase, although this is not immediately obvious, and explicit calculations have to be made for the different energies of a solid/liquid interface. However, the behavior of $\kappa$ strongly suggests that the energy balance is not drastically changed.

Fig.~\ref{fig:kappa3faces} also shows that the rate of increase of disorder is larger in the [110] direction than for the closed packed surfaces (100) and (111). Again this is consistent with the lower formation energy of ordered (111) and (100) surfaces. Another striking difference is that as temperature increases the amount of disorder in the closed packed directions seems to saturate while that for the (110) surface seems to increase approximately linearly with temperature. There is even a slight decrease in $\kappa$ i.e. an increase of ordering, at high temperatures.  Note that data for high temperatures ($T> 1400K$) should be analyzed with caution since at these temperatures there are one or two peaks at most and the fitting procedure is no longer accurate. 

To summarize, the results indicate that the interaction between the ordered solid and the liquid induced layering oscillations within the liquid with a periodicity of the interlayer spacing $d$. If this periodicity is also commensurate with the natural periodicity of the liquid, as given by $2\pi/Q_0$ (see above), then these oscillations are enhanced, and survive a further distance into the liquid. This is the case for the (111) interface. In the case where the d-spacing is very close to $2\pi/Q_0$, such as for the (100) interface, then after some distance (about three layers in Fig.~\ref{fig:interlayer1}) when the effect of the hard-wall has faded, the d-spacing switches to a value close to $2\pi/Q_0$. When the d-spacing is far from $2\pi/Q_0$, then the oscillations fade away after a short distance. 

This suggests that $\kappa$ depends strongly on the d-spacing of the underlying substrate. In fact a reasonable order of magnitude estimate for $\kappa_{\hat{n}}$ as a function of orientation $\hat{n}$, can be obtained by assuming that to first order,  the orientation dependence is related only to the d-spacing by: 
\begin{equation}
 \kappa \sim \displaystyle{1\over d}
\label{eq:fifi2}
\end{equation}
With the value used for the lattice parameter we obtain:
\begin{equation}
\begin{array}{cccc}
\kappa_{111} = 0.433{\rm \AA}^{-1}, & \kappa_{110} = 0.689{\rm
\AA}^{-1}, & {\rm and\;\;} \kappa_{100} = 0.488{\rm \AA^{-1}}.
\end{array}
\end{equation}
Inspection of Fig.~\ref{fig:kappa3faces} shows that this occurs for an intermediate temperature of approximately 1200K. Again we obtain similar $\kappa$ values for the (111) and (100) surfaces, even when neglecting the jump in $d_{100}$ which would make $\kappa$ even more comparable.

The assumption that the layering periodicity within the liquid follows that of the underlying substrate allows us to explain the different rates at which disorder increases with temperature for the different orientations, as shown in Fig.~\ref{fig:kappa3faces}.  If the d-spacing in the liquid region adjacent to the substrate is large, as the case for the closed-packed surfaces (111) and (100), then the overlap between the density peaks is very small, as can be seen from Fig.~\ref{fig:denprof-compare2}.  For open surfaces such as (110) the d-spacing is smaller and the overlap is larger. A small overlap leads to a small number of atoms in between the peaks, as is clearly observed from the deep valleys in Fig.~\ref{fig:denprof-compare2}, while a large overlap leads to a higher density of atoms in between the peaks. In a metallic system, the energy of an atom is lower when it is in a dense environment, and as a result atoms tend to reside in a dense layer rather than in a less dense valley. Breaking the order of the layers essentially means that more atoms leave the ordered layers and reside between the layers.  Therefore, in orientations with a large d-spacing, where valleys are almost void of atoms, the barrier to such interlayer diffusion is large. So for close-packed surfaces (111) and (100), where order prevails up to higher temperatures, the peaks remain well separated, and hence as temperature increases, there is almost no increase in $\kappa$. For the (110) interface, the larger overlap between the peaks allows for easy interlayer diffusion, and hence $\kappa$ increases readily.

We presume that the link between the d-spacing of the liquid and the substrate is not causal, but is due to the imposition of a {\em parallel} structure on the liquid by the substrate surface. If the liquid atoms under the first peak in the liquid density are strongly correlated with the surface atomic structure, then the fact that our substrate was chosen to have nearly the same atomic density as the liquid ensures that the distance to the second liquid peak is the same as the interplanar spacing in the substrate. It might be interesting to demonstrate this by varying the d spacing of the substrate without changing the density within the lattice planes parallel to the surface. We would not expect this to lead to significant changes in the liquid peak separations. Other extensions of this study would be to introduce a substrate with a different lattice structure to that of the frozen liquid, eg. bcc in the case of liquid aluminium. Finally, the effect of varying the lattice parameter of the substrate, such that it is no longer commensurate with the frozen liquid in its normal crystal structure, may induce some interesting effects for further study.

\section{Summary and Conclusions} 
Atomistic simulations on a model liquid-solid interface have been performed, and the density profile of the liquid phase as a function of temperature and substrate orientation has been studied.  The substrate was composed of aluminium atoms pinned to an fcc lattice structure, while the liquid metal was composed of normal liquid aluminum.  Substantial ordering was found in the liquid near the interface.  The density profiles showed large oscillations corresponding to the layering of the liquid near the interface.  The oscillations dampen gradually within the interfacial region according to $\rho (z) \sim \rm{e}^{-\kappa z}$, until the density of the liquid phase is reached.  The ordering was found to extend further into the liquid region at close-packed interfaces such as (111) and (100). Moreover, the periodicity of the liquid layering was found to be strongly correlated to the inter-planar spacing of the rigid substrate, which in turn is related to the planar structure of the underlying substrate. 

 The decay parameter $\kappa$ was defined as a disorder parameter which gives the extent at which the layering extends into the liquid. It was found that $\kappa$ is consistently larger for the (110) interface, than for the (111) and (100) interfaces. Moreover, as temperature increases, $\kappa$ for the open-surface (110) increases approximately linearly with temperature, while for the (111) and (100) interfaces it saturates.  The interlayer spacing in the liquid also determined the amount of overlap between the density peaks. The smaller the d-spacing, the larger the overlap between the ordered liquid layers, which leads to an enhanced interlayer diffusion resulting in a faster increase of disorder.

\bigskip
{\bf Acknowledgements:}
We are grateful to D.G. Brandon for critical review of the manuscript
and valuable suggestions. The work of A.H. was supported by both the
Milton and Edwards Fellowship, and the Israel Science Ministry. We
also acknowledge S.G. Lipson for helpful comments. These calculations
were made on the computers of the Computational Physics Group at the
Technion, the Technion Computer Center, and S. Brandon's Linux
cluster. Support of the Israel Science Ministry under grant 1560199,
the German-Israel Science Foundation under grant 653-181.14/1999, and
the US-Israel Binational Science Foundation (BSF Grant 1998102) is
acknowledged.

\newpage
\begin{itemize}
\item[Figure 1:]  An example of a cell for the simulation of a system with a free {(}110{)} surface. The total number of atoms is \( 1944\), out of which \( 432\) atoms are 'solid' (light gray) and the rest are ``liquid'' (dark gray) aluminium atoms.

\item[Figure 2:] An example of an annealing schedule for system number 5 (see Table\nobreakspace {}1\hbox {}). The state of the system at the final (target) temperature is compared with the initial state. The schedule is repeated until the system is in equilibrium, that is when the energy and density profile in the final state and initial state are the same.

\item[Figure 3:]  The density profile $\rho (z)$ as a function of distance from the hard wall $z$, is shown for a (111) system at two temperatures: $T_1=1000K$, and $T_2 = 1200K$. The insert is a plot of the average number of atoms per bin ${\left \delimiter "426830A N_x\right \delimiter "526930B }$ for the region near to the rigid wall. Data is from a simulation on system number 4 described in Table\nobreakspace {}1\hbox {}.

\item[Figure 4:]  The density profile $\rho (z)$ as a function of distance from the hard wall $z$, is shown for a (110) system at two temperatures: $T_1=1000K$, and $T_2 = 1200K$. The insert is a plot of the average number of atoms per bin ${\left \delimiter "426830A N_x\right \delimiter "526930B }$ for the region near to the rigid wall. Data is from a simulation on system number 2 described in Table\nobreakspace {}1\hbox {}.

\item[Figure 5:] The density profile $\rho (z)$ as a function of distance from the hard wall $z$, is shown for a (100) system at two temperatures: $T_1=1000K$, and $T_2 = 1200K$. The insert is a plot of the average number of atoms per bin ${\left \delimiter "426830A N_x\right \delimiter "526930B }$ for the region near to the rigid wall. Data is from a simulation on system number 1 described in Table\nobreakspace {}1\hbox {}.

\item[Figure 6:] Normalized fine scale density profile for (a): (111), (b):(110), and (c) (100) interfaces at the same temperature (1000K).

\item[Figure 7:] Layer separation across the samples at T=1000K, the dark symbols are within the rigid substrate, and the open symbols are within the liquid.

\item[Figure 8:]
The envelope of the density profile obtained for a (110) sample at $T=1000K$. System numbers are defined in Table\nobreakspace {}1\hbox {}.

\item[Figure 9:] The disorder parameter $\kappa $ vs. temperature for the three surfaces.

\end{itemize}

\newpage
\begin{table}
\caption{Simulation cell geometry.} 
\label{tab:geometry}
\centering
\begin{tabular}{c|c|l|c|c}

{\small System No.} & {\small Interface plane} & {\small No. of planes in each
direction} & {\small No. of rigid planes}&{\small Number of atoms}  \\  \hline

1  & $(100)$ & $22_{[100]} \x  14_{[010]} \x 14_{[001]}$ & 3 &  $2156$   \\
2  & $(110)$ & $12_{[111]} \x  30_{[110]} \x 36_{[211]}$ & 6 &  $2160$   \\
3  & $(110)$ & $18_{[111]} \x  30_{[110]} \x 48_{[211]}$ & 6 &  $4320$   \\
4  & $(111)$ & $18_{[111]} \x  18_{[110]} \x 36_{[211]}$ & 3 &  $1944$   \\
5  & $(100)$ & $40_{[100]} \x  14_{[010]} \x 14_{[001]}$ & 3 &  $3920$   \\
6  & $(111)$ & $32_{[111]} \x  18_{[110]} \x 36_{[211]}$ & 14&  $3456$   \\
7  & $(110)$ & $12_{[111]} \x  34_{[110]} \x 36_{[211]}$ & 10&  $2448$   \\
\end{tabular}

\end{table}

\begin{table}
\caption{Ideal surface formation energies for aluminium  as calculated with the Ercolessi-Adams potential~\cite{Hansen99} and with LDA~\cite{Stumpf}.  \label{tab:energy}
}
\centering
\begin{tabular}{l|cc|cc} 
{\small System} & {\small EA} {\small [eV/\AA$^2$]} & {\small [eV/atom]} & {\small LDA} {\small [eV/\AA$^2$]} & {\small [eV/atom]]}  \\  \hline
$(111)$ & $0.054$ &  $0.38$ &  $0.07$  & $0.48$   \\
$(100)$ & $0.059$ &  $0.48$ &  $0.071$ & $0.56$   \\
$(110)$ & $0.065$ &  $0.74$ &  $0.08$  & $0.89$    \\
\end{tabular}

\end{table}

\end{document}